\documentclass[12pt,a4paper]{article}
\usepackage[utf8]{inputenc}
\usepackage{amsmath}
\usepackage{amsfonts}
\usepackage{amssymb}
\usepackage[english]{babel}
\usepackage{graphicx}
\usepackage{url}
\usepackage{setspace}
\usepackage{natbib}

\newcommand{\blind}{1}

\newcommand{\twoparen}{{(2)}}
\newcommand{\cov}{\hbox{Cov}}
\newcommand{\var}{\hbox{Var}}
\newcommand{\smpxi}{X_{(i)}}
\newcommand{\smpmomtwoTot}{\bar{X}_T^{(2)}}
\newcommand{\smpmomtwoPart}{\bar{X}_p^{(2)}}
\newcommand{\smptwomomenterror}{\bar{X}_e^{(2)}}
\newcommand{\smpfirstmomenterror}{\bar{X}_e}
\newcommand{\g} {g(\smpfirstmomenterror, \smptwomomenterror)}
\newcommand{\momtwo}{\mu_x^{(2)}}
\newcommand{\tauofsqs}{\tau_x^{(2)}}
\newcommand{\vary}{\sigma_y^2}
\newcommand{\varx}{\sigma_x^2}
\newcommand{\varr}{\sigma_R^2}

\newcommand{\condVarPart}{\hbox{Var}(Y | \vec{X}_T, \vec{X}_p)}
\newcommand{\VecsPart}{ \vec{X}_T, \vec{X}_p}
\newcommand{\vecT}{\vec{X}_T}
\newcommand{\vecP}{\vec{X}_p}
\newcommand{\vecs}{\vec{X}_s}
\newcommand{\condVarPartRealized}{\hbox{Var}(Y | \vec{x}_T^*, \vec{x}_p^*)}

\newtheorem{theorem}{Theorem}[section]

\begin{document}
\def\spacingset#1{\renewcommand{\baselinestretch}%
{#1}\small\normalsize} \spacingset{1}
\if1\blind
{
  \title{\bf  Sample Design for Audit Populations}
  \author{Michelle Norris \thanks{
   Michelle Norris is Associate Professor, Department of Mathematics and Statistics, California State University, Sacramento, 6000 J Street, Sacramento, CA 95819 (e-mail:  norris@csus.edu).The author thanks Martin Hauser for his careful proofreading of the manuscript and several helpful discussions.}\hspace{.2cm}\\
    Department of Mathematics and Statistics\\
     California State University, Sacramento\\
    }
  \maketitle
} \fi

\if0\blind
{
  \bigskip
  \bigskip
  \bigskip
  \begin{center}
    {\LARGE\bf Sample Design for Audit Populations}
\end{center}
  \medskip
} \fi

\bigskip

\abstract{We develop several tools for the determination of sample size and design for  MediCal audits. This audit setting involves a population of claims for reimbursement by a healthcare provider which need to be reviewed by an auditor to determine the correct amount for each claim. The existing literature regarding sample planning for audits is incomplete and often includes restrictive assumptions.  To fill these gaps, we exploit the special relationship between the known claim amounts and the unknown post-audit amounts. We propose a hypergeometric generative process for audit populations which we use to derive estimators of variances needed for sample size determination. We further develop a criterion for choosing between simple expansion and ratio estimation and an efficient method for determining exact optimal strata breakpoints in populations with repeated values. 
We also derive a variance estimator under a more general ``partial error” model than previous researches have used.  These tools apply more generally to audits where an overstated book/claim amount is the primary concern and estimation of the total dollar value of the claim errors is the goal.  The sample design methods we develop are illustrated on two simulated audit populations.}


\newpage
\spacingset{1.45} 

\section{Background and Motivation} \label{Intro}
According to the Medicaid program website (\cite{CentersforMed2018B}),

\begin{quote}
``Medicaid provides health coverage to millions of Americans, including eligible low-income adults, children, pregnant women, elderly adults and people with disabilities...The program is funded jointly by states and the federal government."
\end{quote}
In 2016, \$566 billion in Medicaid payments were disbursed to healthcare providers such as pharmacies, medical offices, and school districts in the US (\cite{CentersforMed2018A}). In California, MediCal is the name for the Medicaid program, and the California State Controller's Office is charged with conducting audits to ensure that MediCal funds paid to organizations conform to the requirements of the MediCal program and are of the appropriate amount.  

In planning a MediCal audit, auditors typically have access to a population of MediCal claims which they are charged with auditing for correctness. For example, if the organization is a medical clinic, a single claim may represent a single visit by a single patient, and the population may contain a million claims from a three-year period. The population may account for tens of millions of dollars in disbursed MediCal payments.  Because a complete examination of all claims is not feasible, auditors typically select a sample of claims, then, based on documentation, determine the appropriate amount of MediCal reimbursement that should have been paid for each claim in the sample. There are three possible outcomes for each sampled/audited claim: 
\begin{enumerate}
\item None of the amount claimed is disallowed, and the entire claimed amount is deemed allowable for reimbursement  (as shown in lines 1 and 5 in Table \ref{auditDataTable}).
\item The entire amount claimed is deemed disallowed, and none is deemed allowable for reimbursement (lines 3 and 6 in Table \ref{auditDataTable}).
\item  A portion of the total amount claimed is deemed disallowed and only the remaining portion is allowable for reimbursement(lines 2 and 4 in Table \ref{auditDataTable}). This  case is also called a partial payment or partial error. 
\end{enumerate}
While the claim amounts are known for the entire population prior to an audit, the disallowed amounts are only known \textit{after} the audit and only for the \textit{sampled} claims. We will use both the terms `disallowed amount' and `error amount' to refer to the portion of a claim total that is not allowable for reimbursement.

\begin{table}
\centering
\begin{tabular}{ccccc}
\hline 
Line &Patient ID & Date of Service & Claimed Amount & Disallowed/Error Amount  \\ 
 &&&(known for & (only known \\
 &&& entire population) &for sampled claims)\\
\hline
1&33457 & Jan 15, 2017 & \$52.50 & \$0  \\ 
2&31415 & March 10, 2017 & \$78.90 & \$30.00 \\ 
3&44478 & Oct 27, 2016 & \$25.90 & \$25.90 \\
4&67841 & May 5, 2016 & \$105.00 & \$50.00 \\ 
5&55112 & Nov 20, 2016 & \$125.00 & \$0 \\ 
6&98765 & May 1, 2016 & \$66.00 & \$66.00 \\ 
\hline
\end{tabular} 
\caption{Portion of Hypothetical Data for a MediCal Audit}
 \label{auditDataTable}
\end{table}

The total disallowed amount found in the sample is extrapolated from the sample to the population, and the organization is required to pay that amount or some related amount back to the MediCal fund.  Clearly, maintaining a small margin of error in estimating the total disallowed amount is of interest to all parties. Thus, it is important to design audit samples which estimate the total disallowed amount with a reasonable margin of error while minimizing the sample size. In addition, since a pilot sample is typically an inconvenience to the organization being audited, audit samples must frequently be designed with little to no information about the population of disallowed amounts -- making it difficult to determine sample size and efficient stratification. 


To date, there are few results in the literature pertaining to sample planning for audit populations. In this paper, we will aim to fill some of the gaps and reconcile some differing results in the literature on sample planning for audit populations.  Perhaps the most comprehensive treatment on this topic is contained in the book \textit{Statistical Auditing} by  \cite{roberts1978}. In particular, Roberts derives estimates of the population variances needed for sample size determination under both simple expansion and ratio estimation.  His estimates do not require data from pilot samples.  However, they do require estimating the `error rate,' defined as the proportion of claims in the population containing some error amount or disallowed amount. He uses a Bernoulli generative model to derive his estimates.    \cite{king2013} also propose a Bernoulli model to estimate the variance of the disallowed amounts under simple expansion but arrive at a slightly different estimate.  In this paper, we propose a third method of estimating the variance and reconcile the three different estimators.  In addition, since all currently available methods of determining sample size depend on estimating the error rate or the variance of the population of disallowed values, we also propose a method for determining a conservative sample size which is based solely on the claimed values and does not require any information about the population of disallowed values. 

We also consider the question of choosing between the simple expansion and ratio estimators in simple random sampling.  The general advice on p.157 of  \cite{cochran1977} is to use the ratio estimator instead of simple expansion when $X$, the claim amount, and $Y$, the disallowed amount, satisfy:
\begin{eqnarray}
\dfrac{\hbox{Cov}(X, Y)}{\sigma_x \sigma_y}&>& \dfrac{1}{2}\dfrac{(\dfrac{\sigma_x}{\mu_x})}{(\dfrac{\sigma_y}{\mu_y})} \label{ratiocriteria}
\end{eqnarray}
We specialize this inequality to the audit population setting, and derive a formula for the probability that ratio estimation will outperform simple expansion. Since our formula only relies on the error rate and parameters for the claim population, it can be used for sample planning prior to collecting any information about the population of disallowed amounts. 


A common assumption in existing literature on audit sample design is the `all-or-nothing error assumption' which states that the error/disallowed amount in a claim equals the entire claim amount or zero.  The all-or-nothing error assumption precludes the possibility of partial errors but greatly simplifies theoretic calculations.  Realistically, partial errors do occur in some audit populations so generalizing existing results and deriving new results that apply to more general error models is desirable.  We note that  \cite{neter1977} do consider more general error models in their empirical study, but our research has not revealed any theoretic work on sample design under more general error models. The methods proposed by  \cite{roberts1978} for sample size determination under simple expansion and ratio estimation assume all-or-nothing errors. We extend our method for sample size determination to a population generative model which allows for  partial errors where all partial errors are the same proportion of the claim amount. We prove that, under this simple partial error model, the sample sizes determined will be less than or equal to the conservative sample size for an all-or-nothing error population.

For some audit populations, stratification can significantly reduce the sample size needed to attain a desired margin of error  (compared to simple random sampling).  Often, stratification of audit populations is based on the claim amount; the Cum $\sqrt{f}$ Rule of  \cite{dalenius1959} is used to determine strata boundaries; and optimal allocation is used to allocate the total sample to strata (see \cite{neter1977} and  \cite{budd2008}). However, there is no guarantee that this method will produce the optimal stratification since the Cum $\sqrt{f}$ Rule relies on the assumption that the population is uniformly distributed within each stratum.  In the all-or-nothing errors case and assuming that errors are independent of claim amount, even if the strata are formed so that the claim amounts are roughly uniformly distributed in each strata, the disallowed amounts will not be.  In the best case,  the disallowed amounts in a strata will consists of a point mass at zero and a roughly uniform distribution of the amounts for claims that are fully in error. 

In the audit populations we have had the opportunity to work with, we have observed that even though the population may be very large, some values are repeated many times, making the number of distinct values orders of magnitude smaller than the population size as shown in the Table \ref{uniquetable}.  
\begin{table}
\centering
\begin{tabular}{cc}
\hline
Population Size & Number of Unique Values \\
  \hline 
  21,000 & 100 \\  
  77,000 & 500 \\ 
  1,700,000 & 4000 \\ 
  \hline 
  \end{tabular} 
 
\caption{Population Size and Number of Unique Values in MediCal Populations} 
 \label{uniquetable} 
\end{table}
We prove a theorem with states that when stratifying by claim amount, there can be at most four candidates for optimal strata breakpoints within a run of repeated values in an audit population with a fixed error rate.  This makes it more feasible to find true optimal breakpoints through a complete search when the number of strata and number of unique values in a population is not too large and an estimate of the error rate is available.  \cite{king2013} have studied the gain in stratification assuming the audit population follows a gamma distribution.  Assuming all-or-nothing errors, they study the gain under a two-strata sampling plan with proportional allocation under three different models for the probability a claim is in error.  Their findings suggest that the gain from stratification with proportional allocation and optimal breakpoint choice is most marked in high error rate scenarios. 
\section{Notation and Estimators}
We now summarize two common estimators used to extrapolate the total disallowed amount in an audit.  We start with some notation:
\begin{eqnarray}
N &=& \hbox{the population size} \nonumber \\
\{x_1, x_2,...,x_N\}&=& \hbox{the population of known claimed amounts} \nonumber \\
\{y_1, y_2,...,y_N\} &=& \hbox{the population of unknown disallowed/error amounts} \nonumber \\
\tau_x &=& \sum_{i=1}^N x_i \nonumber \\
\tau_y &=& \sum_{i=1}^N y_i\nonumber \\
R &=& \dfrac{\tau_y}{\tau_x}\nonumber \\
\mu_x &=& \dfrac{1}{N} \tau_x \nonumber \\
\mu_y &=& \dfrac{1}{N}\tau_y \nonumber \\
\sigma_x^2 &=& \dfrac{1}{N}\sum_{i=1}^N (x_i-\mu_x)^2 \nonumber \\
\sigma_y^2 &=& \dfrac{1}{N}\sum_{i=1}^N (y_i-\mu_y)^2 \nonumber \\
\end{eqnarray}
\begin{eqnarray}
n &=& \hbox{the sample size}\nonumber \\
\{x_{(1)}, x_{(2)},...,x_{(n)}\} &=& \hbox{a sample random sample of claims without replacement} \nonumber \\
\{y_{(1)}, y_{(2)},...,y_{(n)}\} &=& \hbox{disallowed values corresponding to sampled claims} \nonumber \\
\bar{y}&=&\dfrac{1}{n}\sum_{i=1}^n y_{(i)} \nonumber \\
\bar{x}&=&\dfrac{1}{n}\sum_{i=1}^n x_{(i)} \nonumber \\
\hat{r} &=& \dfrac{\bar{y}}{\bar{x}}\nonumber \\
\hat{\tau}_{se}&=&N\bar{y}= \hbox{the simple expansion estimator of total error amount} \nonumber \\
\hat{\tau}_r&=&\hat{r} \tau_x = \hbox{the ratio estimator of total error amount} \nonumber\\
\sigma_{\hat{\tau}_{se}}^2 &=& N^2 \cdot \dfrac{\sigma_y^2}{n}\cdot \dfrac{N-n}{N-1} \label{varsimpexp} \\
\sigma_R^2 &=& \dfrac{1}{N} \sum_{i=1}^N(y_i-R x_i)^2 \nonumber \\
\sigma_{\hat{\tau}_{r}}^2 &=& N^2 \cdot \dfrac{\sigma_R^2}{n}\cdot \dfrac{N-n}{N-1}\nonumber
\end{eqnarray}

The rest of this paper is organized as follows.  In Section 3, we  give the sample size formula of interest.  In Section 4, we discuss issues pertaining to sample size determination under the simple expansion estimator.  In particular, we propose a generative model for audit populations; propose an estimator of the variance needed for sample size calculations under an all-or-nothing error model and compare it to existing estimators; extend the procedure for estimating variance to a population with partial payments; and prove a computational shortcut for determining optimal strata boundaries.  In Section 5, we consider the ratio estimator.  We start with a criteria for deciding between simple expansion and ratio estimation; continue by developing a estimator for the variance needed in sample determination under ratio estimation; and finish with comments about stratification and partial errors under ratio estimation.  In Section 6, we apply the sample design tools developed in this paper to two audit populations simulated to reflect actual audit populations in the literature.  We offer some concluding remarks and avenues for further research in Section 7.

\section{Sample Size Formula}
Under simple expansion, the $(1-\alpha) \times 100 \%$ large-sample confidence level margin of error of $\hat{\tau}_{se}$ is
\begin{eqnarray}
E=z_{(1-\frac{\alpha}{2})}\cdot \sigma_{\hat{\tau}_{se}}\label{marginSimpExp}
\end{eqnarray}
where $z_p$ denotes the $p$th percentile of the standard normal distribution. 
Substituting equation (\ref{varsimpexp}) into equation (\ref{marginSimpExp}) and solving for $n$, we obtain the following sample size formula under simple expansion
\begin{eqnarray}
n=\dfrac{z_{(1-\frac{\alpha}{2})}^2 \cdot N^3\sigma_y^2}{E^2(N-1)+z_{(1-\frac{\alpha}{2})}^2N^2\sigma_y^2} \label{SampleSizeSimExp}
\end{eqnarray}
The sample size formula will give the sample size required to attain a chosen margin of error and confidence level provided that the variance of disallowed amounts, $\sigma_y^2$, is known.  However, $\sigma_y^2$ is typically not known in the planning stages of an audit. One could obtain an estimate of $\sigma_y^2$ using a pilot sample, but this is an inconvenience to an audited organization since they would have to pull records twice -- once for the pilot sample and again for the actual full audit. In the next section, we propose a generative model for audit populations which permits estimation of $\sigma_y^2$ in cases where the error rate can be approximated.  Under ratio estimation, the sample size formula is equation (\ref{SampleSizeSimExp}) with $\sigma_R^2$ substituted for $\sigma_y^2$.  We propose methods for estimating $\sigma_R^2$ during the planning stages of an audit in Section \ref{predictsigmar}.  

\section{The Model}

To determine variation in the disallowed amounts based on the known claimed amounts, we now introduce a simple model for audit results which assumes that the audit of any single MediCal claim gives an all-or-nothing error -- that is, that the  claim payment is either entirely allowable or entirely disallowable. Thus, payments that are partially allowed would not occur in this simplified case.  Later, we will generalize the results to a partial payment scenario.  This simplified model also assumes that the probability a claimed payment is disallowed is independent of the amount of the claim. We use the following notation:
\begin{eqnarray*}
 \pi &=& \hbox{the population proportion of error claims}, \quad \pi \in \{0, \frac{1}{N}, \frac{2}{N}, ...,\frac{N-1}{N}, 1\}\\
N_e &=& \pi \cdot N = \hbox{the number of error claims in the population}\\
\end{eqnarray*}
$\pi$ represents  the ``error rate" of the audit population as defined in Section \ref{Intro}. We can conceptualize an actual audit population with error rate $\pi$ as having been generated by choosing a simple random sample without replacement (SRSWOR) of size $N_e$ from the $N$ claims in the population and designating these as the errors.  Let the random vector $\vec{X}_s=\{X_{(1)}, X_{(2)}, ...X_{(N_e)}\}$ denote the SRSWOR of size $N_e$ from the population of claimed amounts $U=\{x_1, x_2,...,x_N\}$. An observed value of $\vec{X}_s$ will be denoted with lowercase letters, i.e. $\vec{x}_s=\{x_{(1)}, x_{(2)}, ...x_{(N_e)}\}$. 

Since all errors are all-or-nothing errors, a random variable $Y$ which is useful for estimation of $\sigma_y^2$ is defined as follows:
\begin{eqnarray}
  Y= \left\{
\begin{array}{ll}
      X_{(i)} & \hbox{ with probability } \dfrac{1}{N} \hbox{ for } i=1,2,...N_e \\
      0 & \hbox{ with probability } 1- \pi \\      
\end{array} 
\right. \label{model}
\end{eqnarray}
$Y$ could be interpreted as a draw from a hypothetical audit population where the claimed amounts are fixed, but the group of $N_e$ claims which are in error has yet to be determined. Hence, the erroneous claims are denoted as random variables $\smpxi$. Under the proposed model, probabilities concerning the $N_e$ claims which are chosen to be erroneous follow a hypergeometric distribution as opposed to the Bernoulli Distribution used in  \cite{roberts1978} and  \cite{king2013} and defined in equation (\ref{modelBinom}). We argue that, at the time of an audit, a fixed number of errors already exist in the population, even though it is not known specifically which claims are in error. Hence, if the error rate is known, an audit population is like a coin that has already been tossed a finite number of times and is best modeled by the hypergeometric distributions as opposed to a coin that \textbf{will} be tossed a finite number of times and should be modeled by a Binomial distribution.

\subsection{Three Potential Estimators of $\vary$}
We now identify and compare three potential estimators of $\vary$. To formulate our proposed estimator of $\vary$, we observe that the unknown audit population would be completely known if the realized vector of erroneous claims $\vec{X}_s$ were known. Suppose that $\vec{X}_s=\vec{x}_s^*$ resulted in the audit population, then
\begin{eqnarray}
\mu_y &=& E(Y | \vec{X}_s=\vec{x}_s^*)\nonumber \\
\sigma_y^2 &=&\hbox{Var}(Y | \vec{X}_s=\vec{x}_s^*) \label{condVar}
\end{eqnarray}
But $\vec{x}_s$ is unknown, so we cannot use equation (\ref{condVar}) to determine $\sigma_y^2$.  We propose using the expected value of $\hbox{Var} (Y|\vec{X}_s)$ given in Theorem \ref{ThmCond}c to estimate $\sigma_y^2$ since it minimizes the mean square prediction error. Theorem \ref{ThmCond} is proven in the Supplement to this paper.

\begin{theorem}[Conditional Expected Value and Variance] \label{ThmCond}
Let $\bar{X}_e = \dfrac{1}{N_e} \sum_{i=1}^{N_e} X_{(i)}$, $\bar{X}_e^{(2)} = \dfrac{1}{N_e} \sum_{i=1}^{N_e} X_{(i)}^2$ and $\mu_x^{(2)}=\dfrac{1}{N}\sum_{i=1}^N x_i^2$. Under the model for $Y$ given in equation (\ref{model}):
\begin{enumerate}
\item[a.] $E(Y|\vec{X}_s) = \pi \bar{X}_e$
\item[b.] $\hbox{Var}(Y|\vec{X}_s) =\pi \bar{X}_e^{(2)}-(\pi \bar{X}_e)^2$
\item[c.] $E(\hbox{Var}(Y | \vec{X}_s)) = \pi \mu_x^{(2)}-(\pi \mu_x)^2-\pi(1-\pi)\dfrac{\sigma_x^2}{N-1}$
\end{enumerate}
\end{theorem}
%
%
%

 \cite{roberts1978} derives a slightly different estimator of the variance $\sigma_y^2$ by assuming the following Bernoulli generating process for the audit population. Letting $X_i=$ the $i$th claim in the population, $Y_i=$ the disallowed value of the $i$th claim for $i=1,2,...,N$ and $0\leq \pi \leq 1$, we assume the audit population is generated as follows: 
\begin{eqnarray}
  Y_i= \left\{
\begin{array}{ll}
      X_i & \hbox{ with probability } \pi \\
      0 & \hbox{ with probability } 1- \pi \\
     
\end{array} 
\right. \label{modelBinom}
\end{eqnarray}
Using the notation of this paper, Roberts derives the following estimator of $\sigma_y^2$:
\begin{eqnarray}
\widehat{\sigma_{(R,y)}^2} = E_{\pi}[\dfrac{1}{N}\sum_{i=1}^N Y_i^2 -\dfrac{1}{N^2}(\sum_{i=1}^N Y_i)^2]\nonumber =\pi \mu_x^{(2)}-(\pi \mu_x)^2-\pi(1-\pi)\dfrac{\sigma_x^2+\mu_x^2}{N} \label{Robertsvar}
\end{eqnarray}

Another option is to use the total variance of $Y$, which can be found using iterated expectations, to estimate $\sigma_y^2$. This result is derived in  \cite{king2013} using the same Bernoulli framework as  \cite{roberts1978} and is also shown in the Supplement.

\begin{theorem}[Total Expected Value and Variance]\label{totalexpvar}
Under the model for $Y$ given in equation (\ref{modelBinom}):

\begin{enumerate}
\item[a.] $E(Y)=\pi \mu_x$
\item[b.] $\hbox{Var}(Y) = \pi\cdot\mu_x^{(2)}-(\pi \mu_x)^2$ 
\end{enumerate}
\end{theorem}

%
%

The federal Office of Inspector General's RAT-STATS software  also uses the total variance of 
$Y$ to estimate $\sigma_y^2$ (RAT-STATs Companion Manual, Rev 5/2010, p. 4-9). The total variance, however, represents the variation in $Y$ as the audit population and the sample from it vary.  We would argue, however, that the audit population is fixed but unknown so that including variation due to a varying audit population in our estimation of $\vary$ is not conceptually satisfying.
 
The three potential estimators of $\sigma_y^2$ are related by the following inequality:

\[\widehat{\sigma_{(R,y)}^2} \leq E(\hbox{Var}(Y | \vec{X}_s)) \leq Var(Y)\] 
However, if the population size, $N$ is large relative to $\sigma_x^2$ and $\mu_x^2$, the term $\pi(1-\pi)\dfrac{\sigma_x^2}{N}$ in Theorem \ref{ThmCond}c and the term $\pi(1-\pi)\dfrac{\sigma_x^2+\mu_x^2}{N}$ in $\widehat{\sigma_{(R,y)}^2}$ will both be close to 0 so that all three estimators will be roughly equal.  This is the case in several audit populations we have reviewed. 

In cases where there is a difference in the estimate of $\vary$ under the binomial and hypergeometric models, it could be argued that the binomial model would be easier to implement since it only requires estimation of the \textit{rate} at which the process of preparing claims generates errors, as opposed to the hypergeometric model where an estimate of the exact proportion of errors in the realized audit population is needed. In a more familiar context, this difference is similar to estimating the rate at which the process of tossing a fair coin will generate heads as 0.50 instead of trying to estimate the proportion of heads that will occur in 20 tosses. Admittedly, the binomial model allows for a variety of realized error rates which follow a binomial probability distribution.  The constraint that the realized error rates follow binomial probabilities, however, seems unnecessarily restrictive. When there is uncertainty in the pre-audit estimate of the error rate, we would propose instead that the hypergeometric model be used and sample sizes be calculated under a range of feasible error rates and the largest resulting sample size be used.  The advantage of the hypergeometric model is that sample sizes can be determined under explicit and flexible assumptions about the distribution of potential realized error rates. 
 
\subsubsection{Estimating $\pi$}\label{conservsmp}
The formula for $E(\hbox{Var}(Y | \vec{X}_s))$ in Theorem \ref{ThmCond}c only depends on the known population of claimed amounts and the error rate, $\pi$.  If an estimate of $\pi$ is available from a past survey or a pilot survey, we can estimate $\sigma_y^2$ using Theorem \ref{ThmCond}c., then substitute the result into equation (\ref{SampleSizeSimExp}) to determine the sample size needed to achieve a given margin of error and confidence level.

If an estimate of $\pi$ is not available, we can obtain a conservative sample size by maximizing $h(\pi)=E(\hbox{Var}(Y | \vec{X}_s))=\pi \mu_x^{(2)}-(\pi \mu_x)^2-\pi(1-\pi)\dfrac{\sigma_x^2}{N-1}$ as a function of $\pi$.  Taking the derivative of $h(\pi)$ and setting it equal to 0 gives:

\begin{eqnarray}
h'(\pi) = \mu_x^{(2)}-2\pi \mu_x^2-(1-2\pi)\dfrac{\sigma_x^2}{N-1}=0 \label{derivOfh}
\end{eqnarray}
Solving equation (\ref{derivOfh}), we obtain
\begin{eqnarray}
\pi_{\hbox{crit}} = \dfrac{1}{2}\cdot \dfrac{\momtwo-\dfrac{\sigma_x^2}{N-1}}{\mu_x^2-\dfrac{\sigma_x^2}{N-1}}\approx \dfrac{\momtwo}{2 \mu_x^2} \label{pi.max}
\end{eqnarray}

In order to maximize $h(\pi)$ over $\pi \in [0,1]$, we must check $h(0), h(1)$ and $h(\pi_{\hbox{crit}})$.  Since $h(0)=0$ and $h(
1)=\sigma_x^2$, the maximum value of $h(\pi)$ is $h_{\hbox{max}}=\hbox{max}\{\sigma_x^2,h(\pi_{\hbox{crit}})\}$. The sample size obtained by substituting $h_{\hbox{max}}$ for $\vary$ in equation (\ref{SampleSizeSimExp}) will be the maximum sample size needed for a specified margin of error and confidence level over all possible error rates, $\pi$.

\subsection{Partial Payments}
Thus far, we have considered a model with all-or-nothing errors. We now wish to consider sample size determination under simple expansion when there are partial payments in the population, i.e. only a portion of the amount claimed is deemed allowable and the remaining portion is disallowable.  We will consider a population of claims where any partial payment is a fixed proportion, $q$ with $0<q<1$, of the claimed amount.  Let $\vec{X_p}=\{X_{(1)}, X_{(2)},..., X_{(p)}\}$ represent the vector of claims partially in error and $\vec{X_T}=\{X_{(1)}, X_{(2)},..., X_{(T)}\}$ where $T>p$ represent claims with any error (partial or full).  Thus, $\vec{X_T}\cap \vec{X_p}^C=\{X_{(p+1)}, X_{(p+2)},...,X_{(T)}\}$ consists of the claims totally in error. Now, the random variable $Y$ which represents a random draw from a random claim population with partial payments is defined as:
\begin{eqnarray}
  Y= \left\{
\begin{array}{ll}
      qX_{(i)} & \hbox{ with probability } \dfrac{1}{N} \hbox{ for } i=1,2,...p 
          \vspace{.2cm}\\
      X_{(i)} & \hbox{ with probability } \dfrac{1}{N} \hbox{ for } i=p+1,p+2,...T \\
      0 & \hbox{ with probability } 1-\dfrac{T}{N} \\
      
\end{array} 
\right.\label{modelpartial}
\end{eqnarray}

Letting $\vec{x}_T^*$ and $\vec{x}_p^*$ be the vectors of realized errors and partial errors, we need to estimate $\vary =\condVarPartRealized$ in order to determine a sample size using (\ref{SampleSizeSimExp}). We recommend using $E(\condVarPart)$ to estimate $\condVarPartRealized$. In the Supplement, we show that
\begin{eqnarray}
E_{\VecsPart}(\condVarPart)\nonumber&=&[\pi_T -\pi_p (1-q^2)]\mu_x^{(2)}-[\pi_T -\pi_p (1-q)]^2\mu_x^2 \nonumber \\
&&-\dfrac{\sigma_x^2}{N-1}[\pi_p (1-q))^2(1-\dfrac{p}{T})\nonumber \\
&&+(1-\pi_T)(1-\dfrac{p}{T}(1-q))(\pi_T-\pi_p(1-q))] \label{vartermN} \\
&<& [\pi_T -\pi_p (1-q^2)]\mu_x^{(2)}-[\pi_T -\pi_p (1-q)]^2\mu_x^2 \label{varpartial2}
\end{eqnarray}
where the upper bound in line (\ref{varpartial2}) is obtained by dropping the second term in the right-hand side of equation (\ref{vartermN}) (which is negative). This bound will be sharp if $\frac{\sigma_x^2}{N-1}\approx 0$ since the coefficient of $\frac{\sigma_x^2}{N-1}$ is less than $2$. This condition is frequently true in MediCal audits so the bound in line (\ref{varpartial2}) will be useful for sample size determination provided estimates of $\pi_T, \pi_p,$ and $q$ are available. If estimates of these quantities are not available, we note that 
\begin{eqnarray}
E_{\VecsPart}(\condVarPart)&<&[\pi_T -\pi_p (1-q^2)]\mu_x^{(2)}-[\pi_T -\pi_p (1-q)]^2\mu_x^2\nonumber \\
&<&[\pi_T -\pi_p (1-q^2)]\mu_x^{(2)}-[\pi_T -\pi_p (1-q^2)]^2\mu_x^2 \nonumber \\
&=& \pi^* \mu_x^{(2)}-(\pi^*)^2\mu_x^2  \label{maxvar}\\
\hbox{where}&& \pi^* =\pi_T -\pi_p (1-q^2) \nonumber
\end{eqnarray}

Taking the derivative and setting it equal to 0, we find that (\ref{maxvar}) is maximized at $\pi^*=\dfrac{\mu_x^{(2)}}{2\mu_x^2}$ so that if no information is available about the audit population, a conservative sample size which will certainly be large enough, but may be larger than necessary, can be calculated by estimating $\vary$ with formula (\ref{maxvar}) and $\pi^*$.  This is close to the estimate of the maximum of $\vary$ in the all-or-nothing errors case. (It may be possible to maximize the expression in (\ref{vartermN}) over $\pi_T, \pi_p$, and $q$, but thus far we have not been able to do so.) It remains an open question to obtain a useful estimator of the variance of $Y$ under more general partial error models.

\subsection{Stratified Sampling under Simple Expansion}
It is often the case that stratified sampling reduces the total sample size needed to attain an estimate of $\tau_y$ having a fixed margin of error (compared to simple random sampling). In stratified sampling, an auxiliary variable which is known to be highly correlated with the study variable is used to partition the population into $L$ disjoint groups called strata.  Then, a simple random sample is selected within each stratum, and, using the following notation, $\tau_y$ is estimated using formula (\ref{stratifiedsimpleexpansion}). 
\begin{eqnarray*}
h &=& 1,2,...,L \hbox{ denotes the stratum}\\
N_h &=& \hbox{ the number of population units in stratum } h\\
x_{1h}, x_{2h},...,x_{Nh,h} &=& \hbox{ denotes the claim amounts in stratum } h\\
y_{1h}, y_{2h},...,y_{Nh,h} &=& \hbox{ denotes the error/disallowed amounts in stratum } h\\
n_h &=& \hbox{ the sample size for stratum } h\\
s_h \subseteq  \{1,2,...,N_h\}&& \hbox{ denotes the subscripts of claims in the sample from stratum } h \\
\mu_{yh}&=& \dfrac{1}{N_h}\sum_{i=1}^{N_h} y_{ih}\\
\sigma_{yh}^2&=&\dfrac{1}{N_h}\sum_{i=1}^{N_h}(y_{ih}-\mu_{yh})^2\\
\bar{Y_h}&=&\dfrac{1}{n_h} \sum_{i \in s_h} y_{ih}
\end{eqnarray*}
The stratified simple expansion estimator of $\tau_y$ and its variance are:
\begin{eqnarray}
\hat{\tau}_{se,st} &=& \sum_{h=1}^{L}N_h \bar{Y_h} \label{stratifiedsimpleexpansion}\\
\hbox{Var}(\hat{\tau}_{se,st}) &=& \sum_{h=1}^{L}  N_h^2 \cdot \dfrac{\sigma_{yh}^2} {n_h}\cdot \dfrac{N_h-n_h}{N_h-1}\label{varTauHat} 
\end{eqnarray}
The factor $\dfrac{N_h-n_h}{N_h-1}$ in equation (\ref{varTauHat}) is the finite population correction factor which is needed when sampling without replacement from a finite population and will be close to 1 if $N_h >> n_h$. Thus, ignoring the finite population correction factor,
\[\hbox{Var}(\hat{\tau}_{se,st}) \approx \sum_{h=1}^{L}  N_h^2 \cdot \dfrac{\sigma_{yh}^2} {n_h}\]
The more homogeneous each stratum is relative to the study variable, the smaller the values of $\sigma_{yh}^2$ and the greater the overall sample size reduction due to stratification (for a fixed precision level).

A number of methods for allocating the overall sample to the strata exist.  These include proportional, equal, or optimal allocation.  Optimal allocation is the allocation method that minimizes $\hbox{Var}(\hat{\tau}_{se,st})$ for fixed strata breakpoints and is given in equation (\ref{optimAlloc}). Under optimal allocation and letting $n$ be the overall sample size, the sample size in stratum $h$ and $\hbox{Var}(\hat{\tau}_{se,st})$ are:
\begin{eqnarray}
n_h &=& \dfrac{N_h \sigma_h}{\sum_{h=1}^{L}N_h \sigma_h}n \label{optimAlloc}\\
\hbox{Var}(\hat{\tau}_{se,st}) &\approx& \dfrac{1}{n}  (\sum_{h=1}^{L}  N_h \sigma_{yh})^2 \label{VarOptimAlloc}
\end{eqnarray}

Typically, the auxiliary variable used to construct the strata is numeric. Thus, it is common to define strata by reordering the population based on ascending values of the auxiliary variable then choosing $L-1$ breakpoints for the auxiliary variable to partition the population into $L$ disjoint groups.  In most populations, the choice of breakpoints can have a substantial effect on $\hbox{Var}(\hat{\tau}_{se,st})$. Since the functional relationship between the strata breakpoints and $\hbox{Var}(\hat{\tau}_{se,st})$ is complicated, no simple, general method exists for finding the optimal breakpoints, given $L$, the number of strata.

Since the 1950's several researchers have proposed methods for finding optimal strata breakpoints.  Early work includes the ``cum $\sqrt{f}$ rule" of  \cite{dalenius1959} discussed in Section \ref{Intro}.  \cite{sethi1963} used an iterative, compute-intensive numerical optimization procedure to implement a condition necessary for optimization put forth by  \cite{dalenius1957}.  However, this method can get stuck in local minima and fail to find the breakpoints yielding the global minimum of $\hbox{Var}(\hat{\tau}_{se,st})$.

With advances in computing speed, recent proposals have primarily focused on random search algorithms (\cite{baillargeon2009} and \cite{kozak2004}).
These methods are usually guaranteed to converge to a local minimum, but not necessarily to the global minimum.  In addition, there is no way to know for certain whether the resulting stratification gives a local or global minimum of the objective function, or, if it gives a local minimum, how far ``off" it is from the global minimum. Moreover, it is often not easy to set up the parameters of the iterative algorithm to enable it to run well or at all.

To ensure the global minimum is found, one could, in theory, perform a complete enumeration by calculating the objective function for every possible choice of the $L-1$ breakpoints and choosing the set of breakpoints which gives the minimum of the objective function.  If the population size is $N$, then there are $C_{L-1}^{N-1}$ possible sets of breakpoints, which is of the order $N^{L-1}$ and the number of calculations grows exponentially as the number of strata increases.  The calculation quickly becomes computationally infeasible for anything but the smallest populations. However, MediCal audit populations often contain a substantial number of repeated values. As an example, Table \ref{uniquetable} shows the approximate population size and number of unique values for three MediCal populations.

While strata breakpoints can occur within a run of repeated values in audit populations, we will next prove a theorem which states that if $\sigma_{yh}^2$ is adequately approximated by Theorem \ref{totalexpvar}b, then there will be at most four potential breakpoints within a run where the objective function in (\ref{VarOptimAlloc}) could assume a minimum value. More specifically, we assume there is a block of $n'$ repeats of a value, say $y$, within a listing of the stratification variable.  We use $x_i$'s to denote the values in this list before the $y$'s and $z_i$'s to denote the values after the run of repeated $y$'s, i.e. this list is denoted $x_1, x_2,...,x_N,\underbrace{y,y,...,y}_{n'},z_1,z_2,...,z_M$. We show that, of all possible breakpoints within the run of $y$'s, there are at most four which are candidates for minimizing $\sum_{h=1}^2 N_h \sigma_{yh}$. Since finding the optimal strata breakpoints now reduces down to checking only four candidates for the entire run of repeats, in populations like those in Table \ref{uniquetable}, the compute time required for a complete enumeration of all possible choices of strata breakpoints will be reduced by roughly 1/50 to 1/100 when our theorem is applied.
\begin{theorem}[Minimum Values of $\sum_{h=1}^2 N_h \sigma_{yh}$ within a Run of Repeated Values]
Let $x_1, x_2,...,x_N,\underbrace{y,y,...,y}_{n'},z_1,z_2,...,z_M$ be a set of positive real numbers.  Let $0 \leq k \leq n'$.  Let stratum 1 consist of $x_1,x_2,...,x_N,\underbrace{y,y,...,y}_{k}$ and stratum 2 consist of $z_1,z_2,...,z_M,\underbrace{y,y,...,y}_{n'-k}$. If $\sigma_{yh}^2$ is well-approximated by $\pi \mu_{xh}^{(2)} -(\pi \mu_{xh})^2$ then the two real values of $k$ which minimize the objective function ,$\sum_{h=1}^2 N_h \sigma_{yh}$ are the roots of the quadratic:
\begin{eqnarray}
\dfrac{1}{2}c_3[c_1^2-c_2^2+2c_3(c_4-c_5)]k^2+[-c_1c_2(c_1+c_2)+2c_3(c_1c_4+c_2c_5)]k+(c_1^2c_4-c_2^2c_5)=0\label{quadmins}
\end{eqnarray}
where
\begin{eqnarray}
c_1&=& \tau_x^\twoparen+Ny^2-2\pi\tau_x y \nonumber\\
c_2&=& \tau_z^\twoparen+My^2+2n'y^2-2\pi y(\tau_z+n'y) \nonumber\\
c_3&=& 2y^2(1-\pi) \nonumber \\
c_4 &=&(M+n')(\tau_z^\twoparen +n'y^2)-\pi(\tau_z+n'y)^2 \nonumber \\
c_5&=&N \tau_x^\twoparen -\pi \tau_x^2 \nonumber
\end{eqnarray}
\hbox{and } $\tau_x = \sum_{i=1}^N x_i$, $\tau_x^\twoparen = \sum_{i=1}^N x_i^2$, $\tau_z = \sum_{i=1}^M z_i$ and $\tau_z^\twoparen = \sum_{i=1}^M z_i^2$.
\end{theorem}
To find the integer value of $k$ which minimizes the objective function, we only need to check the two integer values surrounding the each root of equation  (\ref{quadmins}). The proof of this theorem is given in the Supplement.

\section{Ratio Estimation}
In this section, we give a criterion for deciding whether ratio estimation or simple expansion should be used in an audit.  We derive a reasonable estimator of the variance $\sigma_R^2$ needed under ratio estimation.  Although the proposed estimator of $\sigma_R^2$ depends on the error rate $\pi$, we are able to show that $\pi=0.50$ maximizes the estimated value of $\sigma_R^2$ as in the case of estimation of a binomial proportion. Thus, in cases where $\pi$ is unknown, a conservative sample size can be computed. We also comment on stratification and partial errors under ratio estimation.
\subsection{Choosing Between Ratio Estimation and Simple Expansion}
We now derive a method for determining whether ratio estimation or simple expansion will be more efficient for extrapolating data from an audit sample. Rearranging the criteria for choosing between these two estimators given in inequality (\ref{ratiocriteria}) gives
\begin{eqnarray}
\cov(X, Y)-\dfrac{\varx}{2\mu_x}\mu_y>0 \label{ratiocriteria2}
\end{eqnarray}

In the audit setting, it is not difficult to show that 
$\hbox{Cov}(X, Y|\vecs) = \pi \smptwomomenterror -\pi \mu_x \smpfirstmomenterror$ and $\mu_y = \pi \smpfirstmomenterror $. Making these substitutions into inequality (\ref{ratiocriteria2}) and simplifying, we obtain:
\begin{eqnarray}
 g(\smpfirstmomenterror, \smptwomomenterror)&=&\smptwomomenterror - (\mu_x+\dfrac{\varx}{2\mu_x}) \smpfirstmomenterror \nonumber
 \\
&=& \dfrac{1}{N_e} \sum_{i=1}^{N_e}[X_{(i)}^2- (\mu_x+\dfrac{\varx}{2\mu_x}) X_{(i)}] \nonumber \\
&=& \dfrac{1}{N_e} \sum_{i=1}^{N_e}U_{(i)} \quad \hbox{where } U_{(i)}= X_{(i)}^2- (\mu_x+\dfrac{\varx}{2\mu_x}) X_{(i)} \nonumber \\
&=& \bar{U}_e >0 \nonumber \\
 \label{ratiocriteria4}
\end{eqnarray}
The probability that $g(\smpfirstmomenterror, \smptwomomenterror)>0$ will represent our confidence that the ratio estimator will have smaller variance than the simple expansion estimator.  In order to compute this probability, we determine the distribution of $g(\smpfirstmomenterror, \smptwomomenterror)$.  If $N_e$ is large, the sample mean $ \bar{U}_e =g(\smpfirstmomenterror, \smptwomomenterror)$ will be approximately normal by the Central Limit Theorem. In the Supplement, we show that the mean and variance of $\g$ are:
\begin{eqnarray}
E(\g)&=& \dfrac{1}{2}\varx \nonumber\\
\var(\g) &=&(\dfrac{1}{\pi}-1)\dfrac{1}{N-1}(\sigma_x^{(2) 2}+k^2\varx-2k\mu_{12}')\label{varofg}\\
\hbox{where } k&=& \mu_x+\dfrac{\varx}{2\mu_x} \nonumber\\
\hbox{and } \mu_{12}'&=& \dfrac{1}{N} \sum_{i=1}^N (x_i-\mu_x)(x_i^2-\momtwo) \label{muonetwodef}\\
\hbox{and } \sigma_x^{(2) 2} &=& \dfrac{1}{N}\sum_{i=1}^N (x_i^2-\momtwo)^2 \nonumber
\end{eqnarray}
%
Let  $Z$ represent the standard normal variate. If $N_e$ is large,
\begin{eqnarray}
P(g(\smpfirstmomenterror, \smptwomomenterror)>0)
&\approx& P(Z > \dfrac{0-\dfrac{1}{2}\varx}{\sqrt{(\dfrac{1}{\pi}-1)\dfrac{1}{N-1}(\sigma_x^{(2) 2}+k^2\varx-2k\mu_{12}')}}) \label{negnumerator}\\ 
&>& \dfrac{1}{2}\nonumber
\end{eqnarray}
where the last line is true since the numerator of the right side of line (\ref{negnumerator}) is negative.  The last line implies that ratio estimation is always favored to outperform simple expansion in \textit{any} claim population provided we can assume $\bar{U}_e$ is normally distributed. If normality of $\bar{U}_e$ is not reasonable, a Monte Carlo estimate of the probability that inequality (\ref{ratiocriteria4}) is true would give a more accurate estimate of our confidence that ratio estimation will outperform simple expansion.

As noted in Section \ref{Intro}, ratio-estimator-based confidence intervals can fail to attain the nominal confidence level when applied to audit populations even if the standard large-sample criteria for using ratio estimation are met.  The excess zeros and skewness often found in audit populations require one to check normality assumptions under either estimator to ensure nominal confidence levels are likely to be met with the proposed sample size.  This can be done through Monte Carlo simulation under a range of potential error rates prior to starting an audit.

\subsection{Estimating $\sigma_R^2$} \label{predictsigmar}
We propose $E(\varr | \vecs)$ as an estimator of $\varr$ which can be substituted into the sample size formula in equation (\ref{SampleSizeSimExp}) for $\vary$. 
%
%
%
%
In the Supplement to this paper, we show that:
\begin{eqnarray}
E(\varr | \vecs)=\pi(1-\pi)(\momtwo+\dfrac{\tauofsqs}{\tau_x^2}\varx-\dfrac{2\mu_{12}'}{\tau_x})\label{predvarr}
\end{eqnarray}
Note that equation (\ref{predvarr}) depends on the error rate $\pi$.  As in the case of estimating a binomial proportion, (\ref{predvarr}) is maximized if $\pi = \frac{1}{2}$.  If the error rate cannot be estimated beforehand, using  $\pi = \frac{1}{2}$ will yield a sample size which is sufficiently large for any value of $\pi$ but may be larger than necessary.

 \cite{roberts1978} derives the following analog of this estimator under the binomial generative model:
\begin{eqnarray}
E(\varr | \pi)=\pi(1-\pi)\mu_x^{(2)}[1+\dfrac{1}{N}(\dfrac{\sigma_x^2}{\mu_x^2}+\dfrac{4}{1+(\frac{\sigma_x}{\mu_x})^2}-\dfrac{G_1}{\frac{\mu_x}{\sigma_x}[1+(\frac{\mu_x}{\sigma_x})^2]}-5)]\label{robertsratio}\\
\hbox{where } G_1=\dfrac{\sum_{j=1}^N(X_j-\mu_x)^3}{N\sigma^3} \nonumber
\end{eqnarray}

For large audit populations, $N$ and $\tau_x$ will both be large relative to $\mu_x$ and $\sigma_x^2$ so that equations (\ref{robertsratio}) and (\ref{predvarr}) will give, $E(\varr | \pi) \approx E(\varr | \vecs) \approx \pi(1-\pi)\mu_x^{(2)}$.

\subsection{Stratification Under Ratio Estimation}
It is useful to note that optimal stratification under simple expansion by minimizing Formula (\ref{VarOptimAlloc}) depends on the error rate, $\pi$, since $\pi$ cannot be factored out of the estimated value of $\sigma_y^2$. The analogous problem of finding optimal strata breakpoints under ratio estimation is independent of $\pi$ since formula (\ref{predvarr}) only includes $\pi$ in the multiplicative constant $\pi(1-\pi)$.  Thus, the optimal strata breakpoints under ratio estimation will not be incorrect if the estimated value of $\pi$ is incorrect.  Under simple expansion, however, incorrect estimation of $\pi$ can lead a suboptimal choice of strata breakpoints. In addition, the computational shortcut for finding strata breakpoints for populations with a substantial number of repeated values under simple expansion has an analogous result for the ratio estimator.

\subsection{Partial Errors under Ratio Estimation}
Under the partial error model defined in (\ref{modelpartial}), we have derived the expected value of $\varr$.  It is a complicated expression which is not likely to be of practical use since it requires estimates of $\pi_T$ = proportion of all errors (partial or full), $\pi_p$= proportion of partial errors and $q$= the portion of each partial payment in error (assumed to be the same for all partial errors) and is not easily bounded or approximated by a more feasible expression. 

It is possible on a case-by-case basis to use a simulation study to ascertain whether partial errors are likely to cause $\varr$ to increase over the all-or-nothing error model estimator given in equation (\ref{predvarr}).  For the two example populations in Section \ref{examplesection}, we found that the estimated value of 
$\varr$ will be lower under a range of partial error models unless the error rate is in the 0.80 or 0.90 range. 

%

\section{Audit Example}\label{examplesection}
\begin{figure}
\centering
\includegraphics[scale=.5]{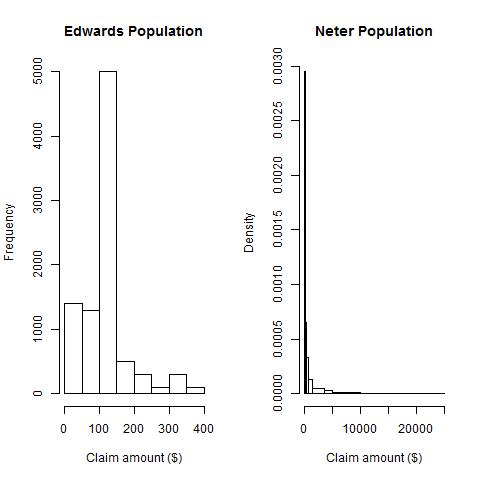}
\caption{The Two Simulated Audit Populations}\label{SimPopfig}
\end{figure}

Since actual MediCal audit data are confidential, we demonstrate these sample design tools using two simulated audit populations. The \textbf{Edwards Population} was simulated to resemble the home health services population in  \cite{edwards2011}.  This population has a low variance and is right skewed with a spike of values in the \$100-150 range.  The population size is 9000, and it represents a paid amount of about \$1.1 million. The \textbf{Neter Population} was simulated to resemble Population 4 on p. 502 of \cite{neter1977}.  This population is also right skewed but with higher variance than the Edwards population.  It contains 4033 items and represents \$7.5 million. Histograms of these populations are shown in Figure  \ref{SimPopfig}.

\subsection{Ratio Estimation versus Simple Expansion}
\begin{figure}
\centering
\includegraphics[scale=.4]{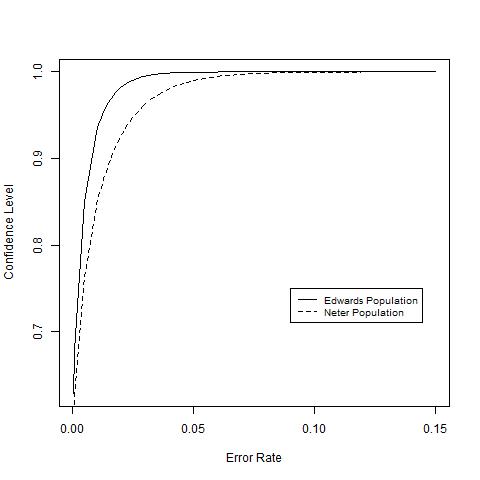} 
\caption{Confidence that Ratio Estimation will Outperform Simple Expansion}\label{ratiovssimplegraph}
\end{figure}

First, we consider the choice between ratio estimation or simple expansion for the example populations. Using the criteria in inequality (\ref{negnumerator}), we can calculate the confidence that ratio estimation will outperform simple expansion over a range of potential error rates.  Figure \ref{ratiovssimplegraph} shows the results of this calculation with a separate graph for each population. Unless error rates are quite low, ratio estimation should be used for either population, assuming the assumptions for ratio estimation hold.

\subsection{Sample Size Under Ratio Estimation}
\begin{figure}
\centering
\includegraphics[scale=.5]{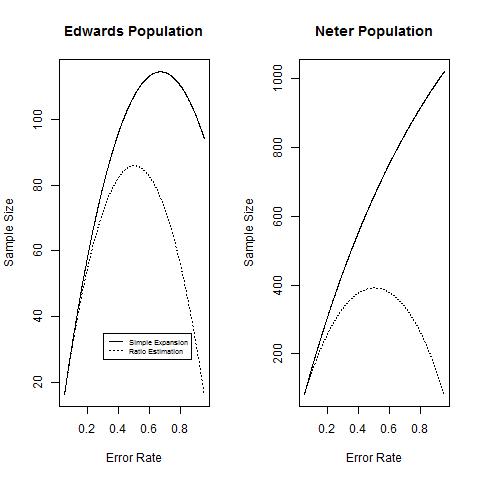}  
\caption{Sample Size as a Function of Error Rate}\label{samplesizefig}
\end{figure}

Suppose that for the Edwards Population representing \$1.1 million in paid claims, we wish to estimate the total error with maximum margin of error \$110,000 (10\% of the total amount paid) at 90\% confidence level.  For the Neter Population, representing \$7.5 million, we wish to estimate the total error with maximum margin of error \$750,000 at 90\% confidence. The sample sizes required over a range of potential error rates is shown for each population in Figure \ref{samplesizefig}. For comparison, the sample size is shown for both estimators even though the ratio estimator is the preferred estimator.
Under ratio estimation, maximal sample sizes occur at an error rate of $\frac{1}{2}$ for any population. This behavior is apparent in Figure \ref{samplesizefig}. However, under simple expansion, the error rate at which the maximal sample size occurs depends on the claim population data.  For the Edwards Population, the error rate at which the maximal sample size occurs is 0.67 using formula (\ref{pi.max}).  For the Neter Population, the error rate yielding maximal sample size is 2.72 which is outside the range of error rates, so $\pi=1$ will give the conservative sample size. 

\subsection{Effect of Partial Errors on Sample Size Determination}
\begin{figure}
\centering
\includegraphics[scale=.6]{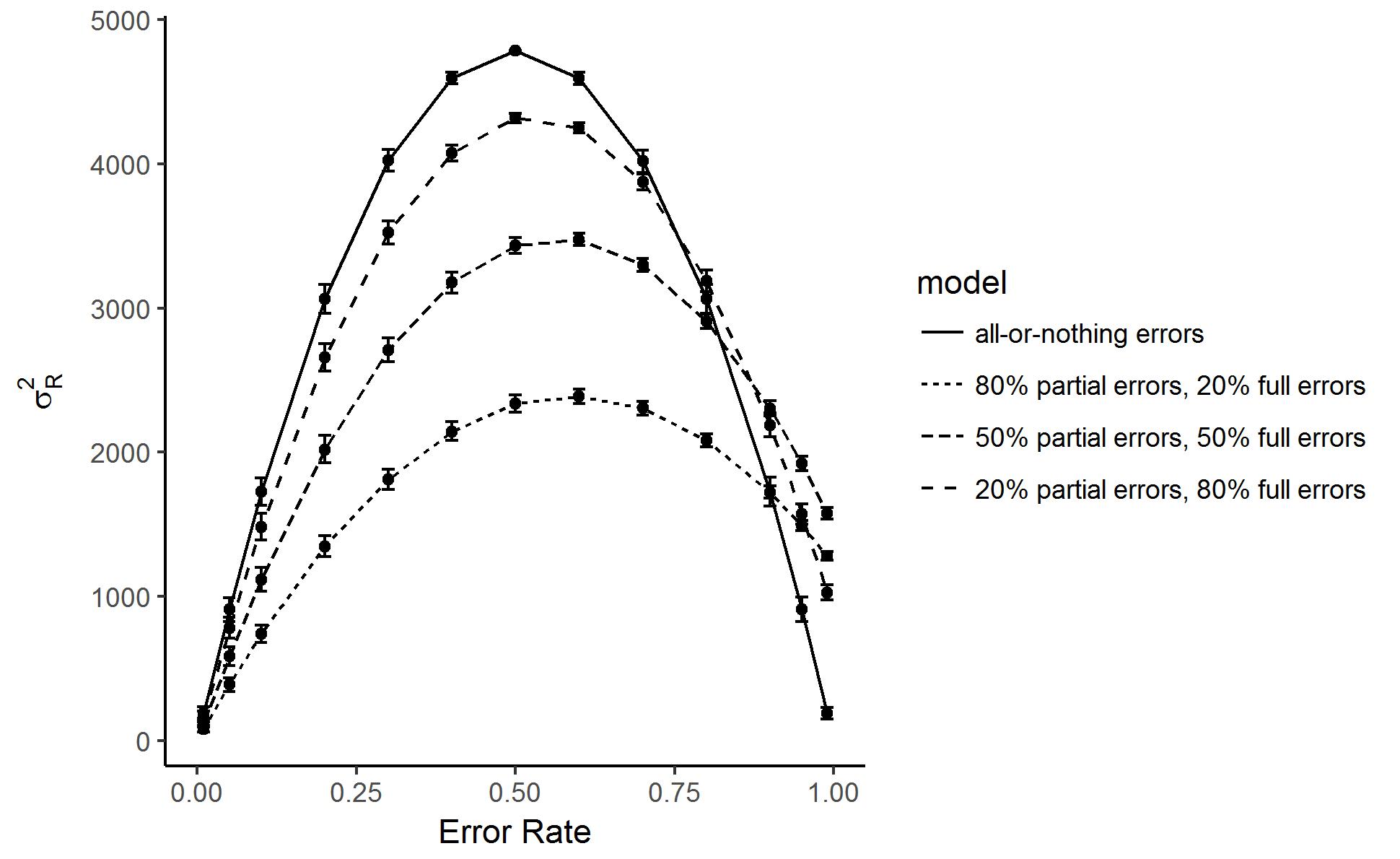}  
\caption{Realized Variances, $\sigma_R^2$, of Simulated Audit Populations from the Edwards Population Under Several Partial Error Models for Various Error Rates (Error bars represent 90\% of the realized variances)}\label{EdwardsPartialErrorFig}
\end{figure}

To determine if partial errors might call for increased sample size under ratio estimation, we conducted a simulation study under a variety of partial error models. For each partial error model, we estimated $\sigma_R^2$ over a range of overall error rates (defined as the proportion of claims fully or partially in error). We simulated four audit populations under the following partial error scenarios:

\begin{itemize}
\item Scenario 1:  All errors are full errors (all-or-nothing errors).
\item Scenario 2:  20\% of errors are full errors, and 80\% are partial errors.
\item Scenario 3:  50\% of errors are full errors, and 50\% are partial errors.
\item Scenario 4:  80\% of errors are full errors, and 20\% are partial errors.
\end{itemize}
To generate the partial errors, the proportion of the claim amount in error was randomly selected from a uniform distribution between 0.2 and 0.8. Figure \ref{EdwardsPartialErrorFig} shows that the realized value of $\varr$ for the Edwards Population will be lower (than the all-or-nothing value) under all partial error models with high probability until the error rate exceeds about 0.80.  The graph for the Neter Population is similar but with larger error bars due to more variability in the original population so we omit it.

\subsection{Stratification of Example Populations}
\begin{figure}
\centering
\includegraphics[scale=0.4]{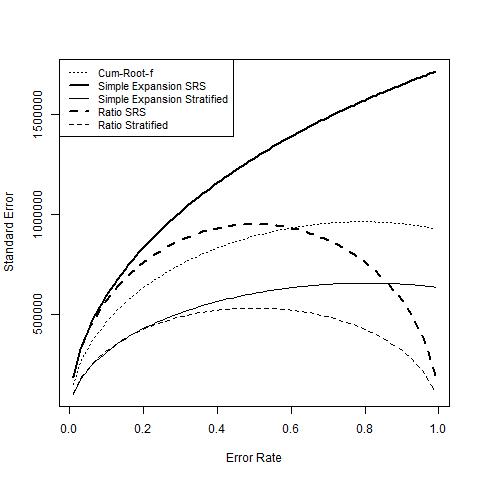}  
\caption{Standard Error For Optimal Two-strata and 
Simple Random Sample (SRS) designs under Simple Expansion and Ratio Esimation as a Function of Error Rate for the Edwards Population}\label{compareStrat}
\end{figure}

We now consider the problem of finding optimal strata breakpoints for the example populations to determine if the precision gain under stratification with two strata is significant. Figure \ref{compareStrat} shows the standard error of the optimal stratification found through an exhaustive search for both the simple expansion and ratio estimators for the Edwards Population. We compare the standard errors across a range of error rates and include the standard error under the Cum $\sqrt{f}$ Rule for comparison.


\section{Conclusions and Further Research}
Using a hypergeometric generative model and assuming all-or-nothing errors, we developed a method for choosing between ratio and simple expansion estimators. We further showed that, for any audit population, ratio estimation is likely to outperform simple expansion as long as the assumptions for ratio-estimator-based confidence intervals are valid. We further derived estimators of the variances needed for sample size calculation under both simple expansion and ratio estimation and found these estimators to give results similar to the binomial-based estimators in the existing literature when the population size is large. Additionally, we developed a computational shortcut for determining exact optimal strata breakpoints in populations with substantial repeats. Notably, these methods can be implemented without pilot study data, requiring only the known claim data and an estimated error rate.  Moreover, in the absence of an estimated error rate, conservative sample sizes can be calculated by maximizing the variance over the error rate, $\pi$. A simple model for partial errors was considered, and we showed that the conservative sample size under the all-or-nothing-error assumption is sufficient under this partial error model. 

A useful avenue for further research would be to extend the results in this paper to more general partial error models which allow the proportion of the claim that is in error to vary and/or which allow for both understatements and overstatements of the claim amount. In addition, since the standard ratio-estimator-based confidence interval may fall short of the nominal confidence level under ratio estimation in audit populations, it would be interesting to investigate whether the estimator of $\varr$ in equation(\ref{predvarr}) would improve the attained confidence level. Finally, since the generative models in this paper all assume that the probability of a claim being in error is independent of the claim amount, it would be useful to rederive these results under a generative model which allows for dependency between the probability of an error and the claim amount.\\

\bigskip
\begin{center}
{\large\bf SUPPLEMENTARY MATERIAL}
\end{center}

\begin{description}

\item[Proofs:] Detailed proofs of theorems in the paper. (ProofsSupplement.pdf)

\item[R Code:] R code for the simulations and calculations in Section \ref{examplesection}.\\ (CodeForSampleDesignForAuditPopulations.R,  functionsForAuditPaperFigs.R)
\end{description}
\vspace{.5in}
\begin{center}
\textbf{\large{Proofs Supplement to \textit{Sample Design for Audit Populations}}}
\end{center}
\setcounter{section}{4}
\begin{theorem}[Conditional Expected Value and Variance] 
Let $\bar{X}_e = \dfrac{1}{N_e} \sum_{i=1}^{N_e} X_{(i)}$, $\bar{X}_e^{(2)} = \dfrac{1}{N_e} \sum_{i=1}^{N_e} X_{(i)}^2$ and $\mu_x^{(2)}=\dfrac{1}{N}\sum_{i=1}^N x_i^2$. Under the model for $Y$ given in equation (\ref{model}):
\begin{enumerate}
\item[a.] $E(Y|\vec{X}_s) = \pi \bar{X}_e$
\item[b.] $\hbox{Var}(Y|\vec{X}_s) =\pi \bar{X}_e^{(2)}-(\pi \bar{X}_e)^2$
\item[c.] $E(\hbox{Var}(Y | \vec{X}_s)) = \pi \mu_x^{(2)}-(\pi \mu_x)^2-\pi(1-\pi)\dfrac{\sigma_x^2}{N-1}$
\end{enumerate}
\end{theorem}

\textbf{Proof:}  
\begin{enumerate}
\item[a.]
\begin{eqnarray*}
 E(Y|\vec{X}_s) &=& \dfrac{1}{N}(\sum_{i=1}^{N_e}\smpxi +(N-N_e)\cdot 0) \\
 &=&\dfrac{N_e}{N}( \dfrac{1}{N_e}(\sum_{i=1}^{N_e}\smpxi)\\
 &=&\pi \bar{X}_e  
 \end{eqnarray*} 
 
 \item[b.] \begin{eqnarray*}
 \hbox{Var}(Y|\vec{X}_s) &=&E(Y^2|\vec{X}_s)-[E(Y|\vec{X}_s)]^2\\
 &=& \dfrac{1}{N}[\sum_{i=1}^{N_e}\smpxi^2 + (N-N_e)\cdot 0^2]-[\pi \bar{X}_e ]^2\\
 &=& \dfrac{N_e}{N}\cdot \dfrac{1}{N_e}[\sum_{i=1}^{N_e}\smpxi^2]- [\pi \bar{X}_e ]^2\\
 &=&\pi \bar{X}_e^{(2)}-(\pi \bar{X}_e)^2
 \end{eqnarray*}
 
 \item[c.] \begin{eqnarray}
 E(\hbox{Var}(Y | \vec{X}_s))&=&\pi E(\bar{X}_e^{(2)})-\pi^2 E(\bar{X}_e^2)\nonumber\\
 &=& \pi \mu_x^{(2)}-\pi^2 [\hbox{Var}(\bar{X}_e)+[E(\bar{X}_e)]^2]  \nonumber\\
 &=&  \pi \mu_x^{(2)}-\pi^2 [\dfrac{\varx}{N_e}(\dfrac{N-N_e}{N-1})+\mu_x^2]\nonumber \\
 &=&  \pi \mu_x^{(2)}-\pi^2\mu_x^2-\pi^2 \dfrac{\varx}{N_e}(\dfrac{N-N_e}{N-1}) \nonumber\\
  &=&  \pi \mu_x^{(2)}-\pi^2\mu_x^2-\pi(1-\pi) \dfrac{\varx}{N-1} \nonumber
 \end{eqnarray}
\end{enumerate}

\begin{theorem}[Total Expected Value and Variance]
Under the model for $Y$ given in equation (\ref{modelBinom}):

\begin{enumerate}
\item[a.] $E(Y)=\pi \mu_x$
\item[b.] $\hbox{Var}(Y) = \pi\cdot\mu_x^{(2)}-(\pi \mu_x)^2$ 
\end{enumerate}
\end{theorem}

\textbf{Proof:} \begin{enumerate}
\item[a.]\begin{eqnarray*}
E(Y)&=&EE(Y|\vecs)\\
&=&E(\pi \bar{X}_e)\\
&=&\pi E(\bar{X}_e)\\
&=&\pi \mu_x
\end{eqnarray*}

\item[b.] Using Theorem \ref{ThmCond}a and c in line two below,
\begin{eqnarray}
\var(Y)&=& E[\var(Y|\vecs)]+\var[E(Y|\vecs)]\nonumber \\
&=& \pi \mu_x^{(2)}-\pi^2 [\hbox{Var}(\bar{X}_e)+[E(\bar{X}_e)]^2] +\var(\pi \bar{X}_e)\nonumber\\
&=& \pi \mu_x^{(2)}-\pi^2 [\hbox{Var}(\bar{X}_e)+[E(\bar{X}_e)]^2] +\pi^2 \var(\bar{X}_e)\nonumber\\
&=&\pi \mu_x^{(2)}-\pi^2 [E(\bar{X}_e)]^2 \nonumber \\
&=&\pi \mu_x^{(2)}-\pi^2\mu_x^2 \nonumber
\end{eqnarray}
\end{enumerate}

\vspace{.5in}

\textbf{Proof of Estimated Variance under Partial Error Model, Equation (\textit{Sample Design for Audit Populations}, Formula (12):}
\begin{eqnarray}
\condVarPart = E(Y^2 | \VecsPart) - [E(Y |\VecsPart)]^2 \label{varcalc}\\
E_{\VecsPart}(\condVarPart) = E_{\VecsPart}(E(Y^2 | \VecsPart)) - E_{\VecsPart}\{[E(Y |\VecsPart)]^2\} \label{iteratedexpectedvar}
 \end{eqnarray}
 
 We first compute the terms in  equation (\ref{varcalc}) .  We use the following notation:  $\bar{X}_T=\dfrac{1}{T}\sum_{i=1}^{T}X_{(i)}$ and $\bar{X}_p=\dfrac{1}{p}\sum_{i=1}^{p}X_{(i)}$.  The total and partial error rates are $\pi_T = \dfrac{T}{N}$ and $\pi_p =\dfrac{p}{N}$. To compute the second term, we calculate
 
 \begin{eqnarray}
 E(Y |\VecsPart)&=& \dfrac{1}{N}(\sum_{i=p+1}^T X_{(i)}+\sum_{i=1}^{p}qX_{(i)}) \nonumber \\
 &=& \dfrac{1}{N}(\sum_{i=1}^{T}X_{(i)}-\sum_{i=1}^{p}X_{(i)}+\sum_{i=1}^{p}qX_{(i)}) \nonumber \\
 &=& \dfrac{1}{N}(\sum_{i=1}^{T}X_{(i)}-(1-q)\sum_{i=1}^{p}X_{(i)})\nonumber \\
  &=& \dfrac{T}{N}\dfrac{\sum_{i=1}^{T}X_{(i)}}{T}-(1-q)\dfrac{p}{N}\dfrac{\sum_{i=1}^{p}X_{(i)})}{p}\nonumber \\
  &=& \pi_T \bar{X}_T-\pi_p (1-q)\bar{X}_p \label{meanpartial}
 \end{eqnarray}
 
 Taking the iterated expectation of (\ref{meanpartial}) and letting $\sigma_{x,T}^2=\dfrac{1}{T}\sum_{i=1}^T (x_{(i)}-\bar{X}_T)^2$, we obtain
 
 \begin{eqnarray}
&&E_{\VecsPart}(E(Y |\VecsPart)^2)\nonumber \\
&=& E_{\vecT}E_{\vecP}[( \pi_T \bar{X}_T-\pi_p (1-q)\bar{X}_p)^2|\vecT] \nonumber \\
 &=& E_{\vecT}{Var}_{\vecP}( \pi_T \bar{X}_T-\pi_p (1-q)\bar{X}_p |\vecT)+E_{\vecT}[E_{\vecP}( \pi_T \bar{X}_T-\pi_p (1-q)\bar{X}_p | \vecT)]^2 \nonumber \\
 &=&E_{\vecT} [\pi_p^2(1-q)^2(\dfrac{T-p}{T-1})\dfrac{\sigma_{x,T}^2}{p}+ \bar{X}_T^2(\pi_T -\pi_p (1-q))^2] \nonumber\\
  &=&\pi_p^2(1-q)^2(1-\dfrac{p}{T})(\dfrac{1}{p})(\dfrac{N}{N-1})\sigma_x^2+ (\dfrac{\sigma_x^2}{T}(\dfrac{N-T}{N-1})\\
  &&+\mu_x^2)(\pi_T -\pi_p (1-q))^2  \label{itersecondterm}
 \end{eqnarray}

 Letting $\smpmomtwoTot =\dfrac{1}{T} \sum_{i=1}^T X_{(i)}^2$ and $\smpmomtwoPart =\dfrac{1}{p} \sum_{i=1}^p X_{(i)}^2$  A similar calculation, gives the first term of equation (\ref{varcalc}):

\begin{eqnarray}
E(Y^2 |\VecsPart)&=& \pi_T \bar{X}_T^{(2)}-\pi_p (1-q^2)\bar{X}_p^{(2)} \label{secondTerm}
\end{eqnarray}

Taking the iterated expectation of (\ref{secondTerm}), we obtain

\begin{eqnarray}
E_{\vecT} E_{\vecP}(Y^2 |\vecT)&=&E_{\vecT}E_{\vecP}( \pi_T \bar{X}_T^{(2)}-\pi_p (1-q^2)\bar{X}_p^{(2)}|\vecT) \nonumber\\
&=&E_{\vecT}( \pi_T \bar{X}_T^{(2)}-\pi_p (1-q^2)\bar{X}_T^{(2)}) \nonumber\\
&=& \pi_T \mu_x^{(2)}-\pi_p (1-q^2)\mu_x^{(2)} \label{iterfirstterm}
\end{eqnarray}

Substituting (\ref{iterfirstterm}) and (\ref{itersecondterm}) into (\ref{iteratedexpectedvar}) and simplifying, we obtain our estimator of $\vary$ when there are partial payments.

\begin{eqnarray}
&&E_{\VecsPart}(\condVarPart)\nonumber \label{varpartial} \\
&=& [\pi_T -\pi_p (1-q^2)]\mu_x^{(2)}  \\
&&-\pi_p^2(1-q)^2(1-\dfrac{p}{T})(\dfrac{1}{p})(\dfrac{N}{N-1})\sigma_x^2- (\dfrac{\sigma_x^2}{T}(\dfrac{N-T}{N-1})\nonumber\\
  &&+\mu_x^2)(\pi_T -\pi_p (1-q))^2 \nonumber
\end{eqnarray}
Rearranging (\ref{varpartial}), we obtain

\begin{eqnarray*}
&&E_{\VecsPart}(\condVarPart)\nonumber \\
&=&[\pi_T -\pi_p (1-q^2)]\mu_x^{(2)}-[\pi_T -\pi_p (1-q)]^2\mu_x^2 \nonumber \\
&&-\dfrac{\sigma_x^2}{N-1}[\pi_p (1-q))^2(1-\dfrac{p}{T})\label{vartermN1} \\
&&+(1-\pi_T)(1-\dfrac{p}{T}(1-q))(\pi_T-\pi_p(1-q))] \label{vartermN} \\
&<& [\pi_T -\pi_p (1-q^2)]\mu_x^{(2)}-[\pi_T -\pi_p (1-q)]^2\mu_x^2 \label{varpartial2}\nonumber
\end{eqnarray*}

\vspace{.5in}
\begin{theorem}[Minimum Values of $\sum_{h=1}^2 N_h \sigma_{yh}$ within a Run of Repeated Values]
Let $x_1, x_2,...,x_N,\underbrace{y,y,...,y}_{n'},z_1,z_2,...,z_M$ be a set of positive real numbers.  Let $0 \leq k \leq n'$.  Let stratum 1 consist of $x_1,x_2,...,x_N,\underbrace{y,y,...,y}_{k}$ and stratum 2 consist of $z_1,z_2,...,z_M,\underbrace{y,y,...,y}_{n'-k}$. If $\sigma_{yh}^2$ is well-approximated by $\pi \mu_{xh}^{(2)} -(\pi \mu_{xh})^2$ then there are at most two real values of $k$ which minimize the objective function ,$\sum_{h=1}^2 N_h \sigma_{yh}$. These critical values can be calculated using claim data and shall be derived in this proof. (To find the $k$ which minimizes the objective function, we only need to check the two integer values surrounding the two critical value $k$.) 
\end{theorem}

\textbf{Proof:}
Let $\tau_x = \sum_{i=1}^N x_i$, $\tau_x^\twoparen = \sum_{i=1}^N x_i^2$, $\tau_z = \sum_{i=1}^M z_i$ and $\tau_z^\twoparen = \sum_{i=1}^M z_i^2$. We wish to minimize the objective function with respect to $k$.
\begin{eqnarray*}
\sum_{h=1}^2 N_h \sigma_{yh}&=& \sqrt{N_1^2 \sigma_{y1}^2}+\sqrt{N_2^2 \sigma_{y2}^2}\\
&=& \sqrt{(N+k)^2[\pi(\dfrac{\tau_x^\twoparen+ky^2}{N+k})-\pi^2(\dfrac{\tau_x+ky}{N+k})^2]}\\
&&+\sqrt{(M+n'-k)^2[\pi(\dfrac{\tau_z^\twoparen+(n'-k)y^2}{M+n'-k})-\pi^2(\dfrac{\tau_z+(n'-k)y}{M+n'-k})^2]}\\
\end{eqnarray*}
Let $h(k)=\sum_{h=1}^2 N_h \sigma_{yh}$. Multiplying out the expressions under the radicals and collecting like terms in $k$, we obtain

\begin{eqnarray}
h(k)&=&\sqrt{\pi}[\sqrt{\dfrac{1}{2}c_3k^2+c_1k+c_5}+\sqrt{\dfrac{1}{2}c_3k^2-c_2k+c_4}\quad]\label{objftn}\\
\hbox{where}&&\nonumber\\
c_1&=& \tau_x^\twoparen+Ny^2-2\pi\tau_x y \nonumber\\
c_2&=& \tau_z^\twoparen+My^2+2n'y^2-2\pi y(\tau_z+n'y) \nonumber\\
c_3&=& 2y^2(1-\pi) \nonumber \\
c_4 &=&(M+n')(\tau_z^\twoparen +n'y^2)-\pi(\tau_z+n'y)^2 \nonumber \\
c_5&=&N \tau_x^\twoparen -\pi \tau_x^2 \nonumber
\end{eqnarray}
Differentiating the right side of (\ref{objftn}) with respect to $k$ and setting it equal to zero gives:
\begin{eqnarray}
h'(k)=\dfrac{c_3k+c_1}{2\sqrt{\dfrac{1}{2}c_3k^2+c_1k+c_5}}+\dfrac{c_3k-c_2}{2\sqrt{\dfrac{1}{2}c_3k^2-c_2k+c_4}}=0 \label{hderiv}\\
\dfrac{c_3k+c_1}{\sqrt{\dfrac{1}{2}c_3k^2+c_1k+c_5}}=\dfrac{-c_3k+c_2}{\sqrt{\dfrac{1}{2}c_3k^2-c_2k+c_4}}\nonumber\\
\dfrac{(c_3k+c_1)^2}{\dfrac{1}{2}c_3k^2+c_1k+c_5}=\dfrac{(-c_3k+c_2)^2}{\dfrac{1}{2}c_3k^2-c_2k+c_4} \label{eqfrac}
\end{eqnarray}
Cross-multiplying (\ref{eqfrac}) and simplifying, the quartic and cubic terms in $k$ conveniently drop out, leaving a quadratic equation in $k$.

\begin{eqnarray}
\dfrac{1}{2}c_3[c_1^2-c_2^2+2c_3(c_4-c_5)]k^2+[-c_1c_2(c_1+c_2)+2c_3(c_1c_4+c_2c_5)]k+(c_1^2c_4-c_2^2c_5)=0\label{quadmins}
\end{eqnarray}
(The roots looks complicated but are quickly calculated on a computer.)
The two roots of Equation (\ref{quadmins}) along with $k=0$ and $k=n'$ are the only possible locations of the minimum of $h(k)$ in the interval $[0,n']$ since $h$ and $h'$ are both defined and continuous on this interval (assuming $\sigma_{yh}\neq 0)$).  We need only evaluate $h(k)$ at the integer values adjacent to each root of (\ref{quadmins}) to find the integer value of $k$ which minimizes $h(k)$. This is because we can show that the expressions under the radicals in equation (\ref{hderiv}) are positive over the domain of $h'$, and, thus, $h'$ can be shown to be continuous on $[0,n']$. Because $h'$ is continuous on $[0,n']$, it must be monotone on the intervals determined by the endpoints and critical points.  Hence, the minimum of $h$ over integer values of $k$ can only occur at integers adjacent to the critical points or at the endpoints of the interval $[0, n']$.

\vspace{.5in}
\textbf{Mean and Variance of Criteria for Choosing Between Simple Expansion and Ratio Esimation, (\textit{Sample Design for Audit Populations}, Formula (22))}
\begin{eqnarray}
E(\g)&=& \dfrac{1}{2}\varx \nonumber\\
\var(\g) &=&(\dfrac{1}{\pi}-1)\dfrac{1}{N-1}(\sigma_x^{(2) 2}+k^2\varx-2k\mu_{12}')\label{varofg}\\
\hbox{where } k&=& \mu_x+\dfrac{\varx}{2\mu_x} \nonumber\\
\hbox{and } \mu_{12}'&=& \dfrac{1}{N} \sum_{i=1}^N (x_i-\mu_x)(x_i^2-\momtwo) \label{muonetwodef}\\
\hbox{and } \sigma_x^{(2) 2} &=& \dfrac{1}{N}\sum_{i=1}^N (x_i^2-\momtwo)^2 \nonumber
\end{eqnarray}

\textbf{Proof of Variance:}
\begin{eqnarray*}
\hbox{Var}( g(\smpfirstmomenterror, \smptwomomenterror)) &=& \hbox{Var}(\bar{U}_e)\\
&=& \dfrac{\hbox{Var}(U_{(i)})}{N_e}\cdot\dfrac{N-N_e}{N-1}\\
&=& \dfrac{\hbox{Var}(X_{(i)}^2)+k^2\hbox{Var}(X_{(i)})-2k\hbox{Cov}(X_{(i)}^2,X_{(i)}) }{N_e}\cdot \dfrac{N-N_e}{N-1}\\
&=& \dfrac{ \sigma_x^{(2) 2}+k^2\sigma_x^2-2k\mu_{12}'}{N_e}\cdot \dfrac{N-N_e}{N-1}\\
&=& \dfrac{ \sigma_x^{(2) 2}+k^2\sigma_x^2-2k\mu_{12}'}{N-1}\cdot (\dfrac{1}{\pi}-1)
\end{eqnarray*}

\vspace{.5in}

\textbf{Proof of Estimator of Variance under Ratio Esimation, (\textit{Sample Design for Audit Populations}, Formula (\ref{predvarr})):}

Let $s = \{(1), (2),...(N_e)\}\subseteq \{1,2,...,N\}\}$, i.e. $s$ contains the subscripts of a random subset of size $N_e$ representing the erroneous claims in the population $\{x_1,x_2,...,x_N\}$.  In this notation, $R=\dfrac{\tau_y}{\tau_x}=\dfrac{\sum_{i \in s}{x_i}+\sum_{i \not{\in} s} 0}{\tau_x}$ so that

\begin{eqnarray}
N\varr | \vecs &=& \sum_{k=1}^N (y_k-Rx_k)^2 \nonumber\\ 
 &=& \sum_{k=1}^N (y_k-\dfrac{\sum_{i \in s}x_i}{\tau_x}x_k)^2\nonumber\\ 
&=&\sum_{k \not{\in} s}(0-\dfrac{\sum_{i \in s}x_i}{\tau_x}x_k)^2+\sum_{k \in s}(x_k-\dfrac{\sum_{i \in s}x_i}{\tau_x}x_k)^2 \label{rawsigmaR}
\end{eqnarray}

To make $E(N\varr | \vecs)$ easier to compute, we multiply out (\ref{rawsigmaR}) and write the resulting expression in terms of the random variables $\smpfirstmomenterror$ and $\smptwomomenterror$.

\begin{eqnarray}
N\varr | \vecs &=& \dfrac{\tauofsqs}{\tau_x^2}N_e^2\smpfirstmomenterror^2+N_e \smptwomomenterror-\dfrac{2N_e^2}{\tau_x}\smpfirstmomenterror \smptwomomenterror \label{varrIntermed}
\end{eqnarray}

Using a result of  \cite{espejo1997}, we find that 

\[E(\smpfirstmomenterror \smptwomomenterror)=(1-\pi)\dfrac{1}{N_e}\mu_{12}'+\mu_x \momtwo\]
 where $\mu_{12}'$ was defined in (\ref{muonetwodef}).  Substituting this last equation into equation (\ref{varrIntermed}) along with the expectations of the remaining random variables then simplifying yields Formula (\ref{predvarr}) in the manuscript.

\bibliographystyle{agsm}

\bibliography{auditrefs}

\end{document}